\documentclass[numberedappendix]{emulateapj}

\usepackage{graphicx}
\usepackage{epstopdf}

\usepackage[center]{subfigure}
\usepackage{float}

\newcommand      \OIII        {O~\textsc{iii}}
\newcommand      \NII        {N~\textsc{ii}}
\newcommand      \HII        {\,H\,{\footnotesize II} } 
\newcommand      \prad     {$P_{\rm rad}$}
\newcommand      \pamb    {$P_{\rm amb}$}
\newcommand      \msun   {M_\odot}
\newcommand      \rch         {\,\tilde{r}_{\rm ch}} 

\begin{document}
\slugcomment{Accepted for publication in ApJ, March 15, 2013}

\title{Line Emission from Radiation-Pressurized \HII Regions.\\II: Dynamics and Population Synthesis}
\author{Silvia Verdolini\altaffilmark{1}, Sherry C. C. Yeh\altaffilmark{2}, Mark R. Krumholz\altaffilmark{3}, Christopher D. Matzner\altaffilmark{2}, Alexander G. G. M. Tielens\altaffilmark{1}}

\altaffiltext{1}{Leiden Observatory, University of Leiden, P. O. Box 9513, 2300 RA Leiden, Netherlands}
\email{verdolini@strw.leidenuniv.nl}

\altaffiltext{2}{Department of Astronomy \& Astrophysics, University of Toronto, 50 St. George St., Toronto, ON M5S 3H4, Canada}

\altaffiltext{3}{Department of Astronomy and Astrophysics, University of California, Santa Cruz, CA 95064 USA}

\begin{abstract}
Optical and infrared emission lines from \HII regions are an important diagnostic used to study galaxies, but interpretation of these lines requires significant modeling of both the internal structure and dynamical evolution of the emitting regions. Most of the models in common use today assume that \HII region dynamics are dominated by the expansion of stellar wind bubbles, and have neglected the contribution of radiation pressure to the dynamics, and in some cases also to the internal structure. However, recent observations of nearby galaxies suggest that neither assumption is justified, motivating us to revisit the question of how \HII region line emission depends on the physics of winds and radiation pressure. In a companion paper we construct models of single \HII regions including and excluding radiation pressure and winds, and in this paper we describe a population synthesis code that uses these models to simulate galactic collections of \HII regions with varying physical parameters. We show that the choice of physical parameters has significant effects on galactic emission line ratios, and that in some cases the line ratios can exceed previously claimed theoretical limits. Our results suggest that the recently-reported offset in line ratio values between high-redshift star-forming galaxies and those in the local universe may be partially explained by the presence of large numbers of radiation pressured-dominated \HII regions within them.
\end{abstract}


\section{Introduction} 

Ratios of optical and infrared lines from \HII regions are popular diagnostics that have been used to infer a large number of properties of galaxies. Perhaps the most famous example of this is the \citet{baldwin81} diagram (hereafter the BPT diagram), which plots [\OIII]$\lambda 5007$/H${\beta}$ versus [\NII]$\lambda 6584$/H${\alpha}$. \HII regions in the local universe form a narrow sequence in this diagram, and their position along this sequence provides information about properties of the \HII region such as its density and metallicity. Recently, thanks to the Sloan Digital Sky Survey (SDSS) \citep{brinchmann04, tremonti04}  the sequence has been extended to unresolved galaxies in the local universe. The SDSS showed that galaxies whose line emission is dominated by an active galactic nucleus (AGN) or by fast shocks are distinguishable in the BPT diagram from those whose emission is powered predominantly by star formation. Star-forming galaxies and \HII regions in the local universe follow the same sequence, suggesting that star forming galaxies can be simplified as a collection of \HII regions. In contrast, AGN-dominated galaxies lie off this sequence.

However, star-forming galaxies at high redshift appear to be offset (upward and to the right) in the BPT diagram from those in the local universe \citep{shapley05, liu08, erb06a, erb10}, but do not occupy the same locus as local AGN-dominated galaxies either. Several possible causes for the offset have been suggested. One possibility is that \HII regions at  $z\approx2$ follow the same star-forming sequence as in the local universe, but the presence of an unresolved AGN or shocked gas contaminates their line emission causing the shift in the BPT diagram (e.g., \citealt{liu08}). Observational support for this idea comes from \citet{wright10}, who demonstrate using integral field spectroscopy that a weak AGN is responsible for the shift of a $z=1.6$ galaxy. \citet{trump11} stack HST grism data from many galaxies to show that this phenomenon is reasonably common. However, another possible explanation for the offset is that there are systematic differences exists between \HII regions in the local universe and at high redshift. This suggests that the time is ripe for a reinvestigation of the physics driving \HII region line emission, and thus the location of galaxies in diagnostic line ratio diagrams such as the BPT plot.
 
The problem of computing the integrated line emission produced by a galaxy containing many \HII regions can be roughly decomposed into two separate steps. The first is determining the internal structure of an \HII region given its large-scale properties, for example the radius of the ionization front and the luminosity of the star cluster that powers it. The second is determining the dynamics of the \HII region population in a galaxy, which sets the distribution of \HII region properties. The first of these problems is generally solved with by a radiative transfer and chemical equilibrium code such as Cloudy \citep{ferland98} or MAPPINGS \citep{sutherland93a, dopita00, kewley01}, while the second is solved by a population synthesis code that generates a population of \HII regions and follows their expansion in the interstellar medium \citep[e.g.][]{dopita06}. For this second step, the results depend on what drives \HII region expansion, i.e.~whether \HII regions are classical Str\"omgren spheres whose expansion is driven by warm gas pressure \citep{spitzer78}, wind bubbles whose expansion is controlled by the pressure of shocked stellar wind gas \citep{castor75a, weaver77a}, radiation pressure-driven shells \citep{krumholz09, murray10a}, or something else.

The most commonly-used population synthesis models, those of \citet{dopita00, dopita05a, dopita06b, dopita06} and \citet{groves08a}, assume that the expansion of \HII regions is primarily wind-driven. However, recent resolved observations of \HII regions in nearby galaxies have shown that this assumption is likely to be incorrect. \citet{harperclark09} and \citet{lopez11} use X-ray observations of Carina and 30 Doradus, respectively, to directly estimate the pressure of the shocked hot gas inside expanding \HII regions.\footnote{Note that \citet{pellegrini11a} analyze the same region (30 Doradus) as \citet{lopez11} and report a much higher pressure in the X-ray emitting gas, such that this pressure exceeds radiation pressure. They reach this result by adopting a small filling factor for the X-ray emitting gas, compared to \citeauthor{lopez11}'s assumption of a filling factor close to unity. However, with such a small filling factor, the hot gas is not dynamically important for the \HII region as a whole, and thus the general conclusion that hot gas is dynamically unimportant remains true even if \citeauthor{pellegrini11a}'s preferred filling factor is correct.}  By comparing these pressures to the other sources of pressure driving the expansion, and to the values expected for a wind bubble solution, they conclude that the giant \HII regions cannot be expanding primarily due to shocked wind gas pressure, and that radiation pressure may well be dominant. Moreover, \citet{yeh12} found no evidence for wind-dominated bubbles either in individual regions or on galactic scales, using observed ionization parameters. Physically, the surprisingly weak role of winds is likely a result of \HII regions being ``leaky", so that the hot gas either physically escapes, or it mixes with cooler gas, and this mixing cools it enough for radiative losses to become efficient \citep{townsley03a}. Regardless of the underlying cause, though, the observations clearly show that the wind bubble model should be reconsidered.

In this work, we investigate the implications of these observations, and more broadly of varying the physics governing \HII regions expansion, for line emission and line ratio diagnostics. To do so we create a population synthesis model in the spirit of Dopita's work, and within this model we systematically add and remove the effects of radiation pressure, and we vary the stellar wind strength. In a companion paper (\citealt{yeh12a}, hereafter Paper I) we generate a series of hydrostatic equilibrium models of \HII regions using Starburst99 \citep{sb99} and Cloudy \citep{ferland98}, both including and excluding radiation pressure and stellar winds. In this paper we use these models to predict the integrated line emission of galaxies containing many \HII regions. 

The structure of the remainder of this paper is as follows. In \S~\ref{S:method} we describe the method we implement to generate synthetic galaxies. In \S~\ref{S:results} we analyze the main results, with particular attention to how various physical mechanisms affect observed line ratios, and in \S~\ref{S:obs} we compare to observations. We finish with discussion and conclusions in \S~\ref{S:conclusions}.

\section{Method}\label{S:method}

We are interested in the computing the total line emission of multiple \HII regions in an unresolved galaxy, such those at high redshift, in order to create a synthetic set of data that is directly comparable with observed galaxies in the BPT diagram or similar line ratio diagrams. The procedure consists of two parts. First, we create synthetic line emission predictions for a variety of single \HII regions over a large grid in stellar luminosity, radius, and age. We describe this procedure in detail in Paper I, but for convenience we briefly summarize it below. Second, we build a population synthesis code that creates, evolves and destroys \HII regions, and computes the summed line emission.

  \subsection{Spectral synthesis and photoionization models} \label{SS:method_1}
We create a population of static, single \HII regions, with a wide range of sizes and ionizing luminosities. To do so we use Starburst99 \citep{sb99} to generate ionizing continua from coeval star clusters of different ages. We feed the synthetic spectra into Cloudy 08.00, last described by \citet{ferland98}, as the ionizing continuum emitted at the center of each simulated \HII region. Each \HII region is spherical and in perfect force balance. We adopt Cloudy's default solar abundances and ISM dust grain size distributions, and the same gas phase abundances as \cite{dopita00}. We compute a grid of models covering a wide range in density, from $\log n_{\rm H,in} = -1$ to 5, where $n_{\rm H,in}$ is the number density of hydrogen nuclei at the inner boundary of each \HII region. Each set of the simulations outputs the integrated luminosity of selected optical emission lines, including H$\alpha$, H$\beta$, [\OIII]$\lambda5007$, and [\NII]$\lambda6584$, i.e.~the lines that enter the BPT diagram. For more details we refer readers to Paper I.

\begin{deluxetable}{lcc}
\tabletypesize{\scriptsize}
\tablewidth{0pt}
\tablecaption{Static \HII region Models}
\tablehead{ \colhead{Model} & \colhead{\prad} & $\log\Omega$ }
\startdata
RPWW (Radiation Pressure Weak Winds) & yes & $-1.5$ \\
RPSW (Radiation Pressure Strong Winds) & yes & $2$ \\
GPWW (Gas Pressure Weak Winds) & no & $-1.5$ \\
GPSW (Gas Pressure Strong Winds) & no & $2$
\enddata
\label{tab:models}
\end{deluxetable} 

We compute four sets of static \HII region models, corresponding to four combinations of radiation pressure (\prad) and stellar wind strength (Table \ref{tab:models}). In the models with \prad, radiation pressure is allowed to exceed ionized gas pressure, in contrast to Cloudy's default setting. For models where radiation pressure is absent, the outward force due to the incident radiation field is turned off. We parameterize the strength of the stellar wind by $\Omega$, which is defined as 
\begin{equation}
\Omega \equiv \frac{P_{\rm in}V_{\rm in}}{P_{\rm IF}V_{\rm IF} - P_{\rm in}V_{\rm in}},
\label{EQ:Omega} 
\end{equation}
where $P_{\rm IF}V_{\rm IF} - P_{\rm in}V_{\rm in}$ is the difference of the product of gas pressure and volume between the ionization front ($P_{\rm IF}V_{\rm IF}$) and the inner edge of the \HII region ($P_{\rm in}V_{\rm in}$), which is the outer edge of a hot, wind-pressurized bubble. $\Omega$ is the same wind parameter defined in \cite{yeh12}, and we refer readers to Table 1 and Section 4.1 in that paper for detailed discussion of its meaning. However, an intuitive explanation of $\Omega$ is that it measures the relative energy content of the hot stellar wind gas and the warm photoionized gas; high values of $\Omega$ correspond to wind-dominated \HII regions, while low values to ones where winds are dynamically unimportant.

  \subsection{Population synthesis code}
  \label{sec:synthesis}

We treat a galaxy as a collection of \HII regions only, with no contribution to line emission from other sources (e.g.~stars or warm ionized medium).  We generate, evolve, and destroy these \HII regions using a population synthesis code derived from the \textsc{gmcevol} code described in \citet{krumholz06} and \citet{goldbaum11a}. In our models, we characterize a galaxy by two parameters: a (constant) star formation rate (SFR) and a mean ambient pressure $P_{\rm amb}$, and we give fiducial values of these parameters in Table \ref{tab:fiducial}, though below we explore how our results depend on these choices. For all the results described in this paper, we run our simulation code for 200 Myr, and write output every 1 Myr. We describe each step the code takes below.

\begin{deluxetable}{lc}
\tablewidth{0pt}
\tablecaption{Fiducial parameters}
\tablehead{
\colhead{Parameter} &
\colhead{Value}
}
\startdata
M$_{a,\rm min}$ 	& 20 M$_{\sun}$ \\
M$_{a,\rm max}$ 	& 5$\times 10^9$ M$_{\sun}$ \\
$k_{\rho}$	& 1 \\
$\mathcal{M}$ & 30 \\
$P_{\rm amb}/k_B$		& $10^4$ K cm$^{-3}$ \\
SFR		& 1 M$_{\sun}$ yr$^{-1}$ \\
\prad 	& yes \\
$f_{\rm trap}$ 	& 2 \\ 
$\phi$ 	& 0.73 \\
$\psi$ 	& 3.2
\enddata
\label{tab:fiducial}
\end{deluxetable}

\paragraph{Creation} To create \HII regions, we pick a series of stellar association masses $M_a$ from a probability distribution 
\begin{equation}  
 p(M_a) \propto M_a^{-2} 
\end{equation}  
in the range $M_{a,\rm min}$ to $M_{a,\rm max}$ \citep{williams97}. We give fiducial values of the minimum and maximum masses in Table \ref{tab:fiducial}, but experimentation shows that these choices have almost no effect on our final result. Each association appears at a time dictated by the SFR; for example, if the first three associations drawn in a calculation have masses of $10^5$ $\msun$, $10^6$ $\msun$, and $10^7$ $\msun$, and the SFR is 1 $\msun$ yr$^{-1}$, the first association turns on at $0.1$ Myr into the simulation, the second at 1.1 Myr, and the third at 11.1 Myr. When an association turns on, we pick stars from a \citet{kroupa01} IMF until we have enough stellar mass to add up to the association mass. For computational convenience we discard stars with masses below 5 $M_{\odot}$, since these contribute negligibly to the ionizing luminosity. For the stars we retain, we use the fits of \cite{parravano03} to assign an ionizing luminosity and a main sequence lifetime. Each association becomes the power source for a new \HII region, with an ionizing luminosity determined by the sum of the ionizing luminosities of the constituent stars. Note that we account for aging of the stellar population in the ionizing spectrum, but use step-function approximations for the luminosity, ionizing luminosity and wind trapping factor in our dynamical calculations.  These we take to be constant during the ionizing lifetime of each cluster.

\paragraph{Expansion} 
The neutral gas in which each \HII region expands has a radial density profile $\rho = \rho_0(r/r_{0})^{-k_{\rho}}$, and our code allows $k_\rho = 0$ or 1. As we discuss in Section \ref{SS:EnvPara}, this choice proves to make very little difference, so unless stated otherwise we simply adopt $k_\rho = 1$. We determine the mean values of $\rho_0$ and $r_0$ from two constraints, one related to the pressure of the galaxy and a second from the mass of the association. Specifically, we require that
\begin{eqnarray}
\label{eq:rho0_1}
M_a & = & [4\pi/(3-k_\rho)] \bar{\rho}_0 \bar{r}_0^3 \\
P_{\rm amb} & = & 2\pi G \left(\bar{\rho}_0 \bar{r}_0\right)^2
\label{eq:rho0_2}
\end{eqnarray}
The first of these equations is equivalent to the statement that the mass of the association is comparable to the mass of the surrounding gas (i.e.~that the star formation efficiency in the vicinity of an association is $\sim 50\%$), while the second is equivalent to the statement that the gas around an association is in approximate pressure balance with the mean pressure of the galaxy. These two statements uniquely determine $\bar{\rho}_0 \bar{r}_0$, but we add a random scatter on top of this to represent the expected density variation present in a turbulent medium. Such media have density distributions well-described by lognormal distributions \citep[e.g.][]{padoan02a}. We therefore scale our value of $\rho_0$ by a factor $x$ drawn from the distribution
\begin{equation}
\label{eq:lognormal}
p(x) = \frac{1}{\sqrt{2\pi \sigma_x^2}} \exp\left[-\frac{\left(\ln x - \overline{\ln x}\right)^2}{2\sigma_x^2}\right],
\end{equation}
where $\overline{\ln x} = \sigma_x^2 / 2$, the dispersion of pressures is $\sigma_x = \sqrt{\ln (1 + \mathcal{M}^2/4)}$, and $\mathcal{M}$ is the Mach number that characterizes the turbulence.  Thus the final value of $\rho_0 r_0$ we adopt for a given \HII region is $\bar{\rho}_0 \bar{r_0} x$, with $\bar{\rho}_0 \bar{r}_0$ determined by the solution to equations (\ref{eq:rho0_1}) and (\ref{eq:rho0_2}), and $x$ chosen from the distribution given by equation (\ref{eq:lognormal}). We adopt a fiducial Mach number $\mathcal{M} = 30$, appropriate for giant molecular clouds in nearby galaxies, but we have experimented with values up to $\mathcal{M} = 300$, appropriate for ultra luminous infrared galaxies (see \citealt{krumholz07g} for more detailed discussion). We find that the choice of $\mathcal{M}$ makes little difference to the final result.

Once we have the density distribution around an \HII region, we can compute its expansion. We do so in two possible ways. The first is simply following the classical \citet{spitzer78} similarity solution for gas pressure-driven expansion, generalized to our density profile. The second is using the \citet{krumholz09} generalization of this solution to the case where radiation pressure is dynamically significant. For this case, we use the approximate solution given by equation (13) of \citeauthor{krumholz09}. This solution involves a few free parameters, and the values we adopt are summarized in Table \ref{tab:fiducial}. The most important of these is $f_{\rm trap}$, which represents the factor by which trapping of photons and wind energy within the expanding dust shell amplifies the radiation pressure force. We adopt a relatively low value $f_{\rm trap} = 2$ as a fiducial value, based in part on recent simulations indicating the radiative trapping is likely to be very inefficient \citep{krumholz12c, krumholz12f}, but we also explore different values of $f_{\rm trap}$ below. Note that in the case $f_{\rm trap} = 0$, the \citet{krumholz09} solution reduces to the classical \citet{spitzer78} one. We discuss the remaining free parameters below. Finally, note that we do not consider the case of expansion following a \citet{weaver77a} wind bubble solution, both because \citet{dopita06} have already obtained results in this case, and because the observations discussed in the Introduction suggest that this model is unlikely to be correct.
  
\paragraph{Stalling} We stop the expansion of an \HII region if its internal pressure ever falls to the pressure of the ambient medium (\pamb). We can express the internal pressure as the sum of the thermal pressure of the ionized gas and the radiation pressure. The thermal pressure of the ionized gas $P_{\rm gas}$ is
\begin{equation} \label{EQ:Pgas} 
P_{\rm gas} = \mu_{\rm H} n_{\rm II} m_{\rm H} c_{\rm II}^2 
\end{equation}  
where $n_{\rm II}$ is the number density of hydrogen nuclei in the \HII region, $c_{\rm II} = 9.74$ km s$^{-1}$ is the sound speed, $\mu_{\rm H}=1/X=1.33$ is the mean mass per H nucleus in units of amu, and $X=0.75$ is the hydrogen mass fraction. We derive $n_{\rm II}$ from photoionization balance, which requires that
\begin{equation}  
\phi S =  \frac{4}{3} \pi r^3 \alpha_B n_{\rm II} n_e = \frac{4}{3} \pi r^3 \alpha_B  \left(1 + \frac{Y}{4X}\right) n_{\rm II}^2
\end{equation}  
where $S$ is the number of ionizing photons per second injected into the region, $n_{\rm H}$ is the number density of H nuclei, $n_e$ is the number density of electrons and $\alpha_B$ is the case-B recombination coefficient. The factor $1 + Y/4X = 1.1$  (assuming helium mass fraction $Y = 0.23$, and that He is singly ionized) accounts for the fact that there are electrons from He as well as from H, and the factor of $\phi = 0.73$ accounts for ionizing photons that are absorbed by dust instead of hydrogen. Thus we have
\begin{equation}  \label{EQ:nII} 
n_{\rm II} = \sqrt{\frac{3 \phi S} {4 \pi r^3 \alpha_B (1 + \frac{Y}{4X})}}.
\end{equation}  
Note that this expression implicitly assumes that the density within the \HII region is constant, which is not the case if radiation pressure exceeds gas pressure. However, in this case the gas pressure is non-dominant, so it matters little if we make an error in computing it. The radiation pressure $P_{\rm rad}$ is
\begin{equation}  \label{EQ:Prad} 
P_{\rm rad} = \frac{\psi S \epsilon_0 f_{\rm trap}}{4 \pi r^2 c}
\end{equation}  
where $\psi=L/(S \epsilon_0)$ is the ratio of the star's bolometric power to its ionizing power counting only an energy $\epsilon_0=13.6$ eV per ionizing photon. We adopt $\psi = 3.2$ following \citet{murray10}, \citet{fall10}, and \citet{lopez11}.

\paragraph{Destruction} We remove an \HII region from our calculation when the stars that provide half its total ionizing luminosity reach the end of their main sequence lifetimes. This may occur before or after stalling, depending on the ambient conditions.

\subsection{Calculation of the line emission}

The population synthesis code generates output files containing information about the \HII regions present at each timestep. For each \HII region, we keep track of the ionizing luminosity $S$ of the driving stellar association, the radius $R_{\rm IF}$ of the ionization front, and the age $t$ of the association. In order to assign line emission luminosities to each \HII region, we perform a three-dimensional interpolation on $R_{\rm IF}$, $S$, and $t$, using the tables of individual \HII region models described in Section \ref{SS:method_1}.

  \begin{figure}
    \centering
%    \epsscale{0.8}
    \plotone{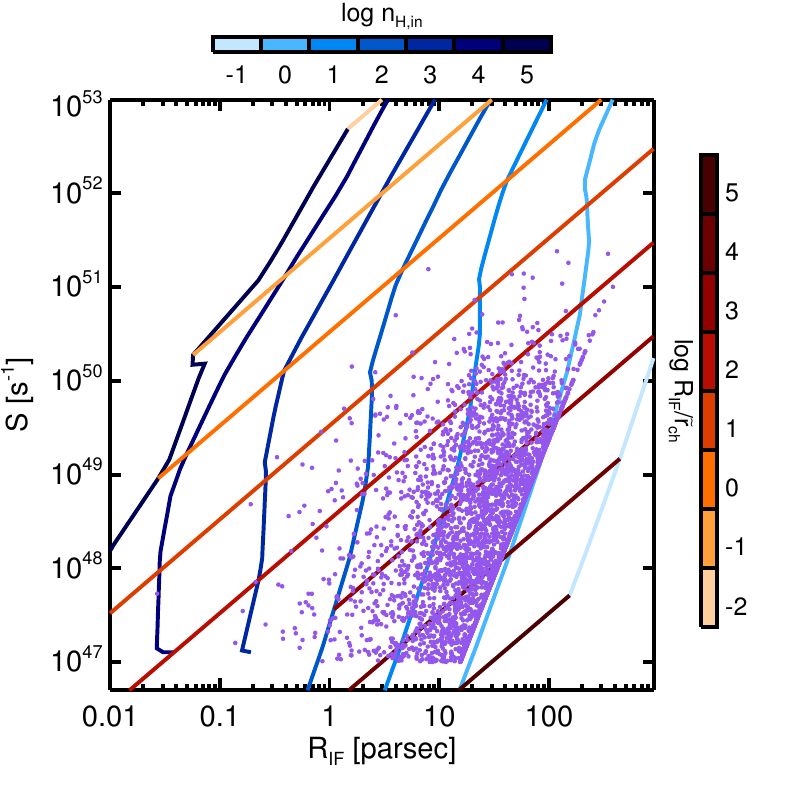}
    \caption{Radius of the ionization front ($R_{\rm IF}$) versus ionizing luminosity ($S$). Each data point represents a single \HII region at one time step of one of our population synthesis calculations; the line against which points have accumulated on the right side of the plot is the stalling line (see Section \ref{sec:synthesis}). Colored lines indicate radii and ionizing luminosities of the \HII regions in the RPWW model grid (see Section \ref{SS:method_1}); note that only a subset of the models are shown in order to avoid clutter. Blue colors indicate models with constant density $\log n_{\rm H,in}$, and red colors indicate models of constant $\log R_{\rm IF}/\rch$, where $\rch$ is the characteristic radius at which radiation and gas pressure balance \cite{yeh12}; note that $\rch$ is a function of $S$ alone, and does not depend on $R_{\rm IF}$. The values of $\log n_{\rm H, in}$ and $\log R_{\rm IF}/ \rch$ are as indicated in the color bars. }
    \label{RIF_S_SFR1_fiducial}
  \end{figure}

\begin{figure}
    \centering
%     \epsscale{0.8}
   \plotone{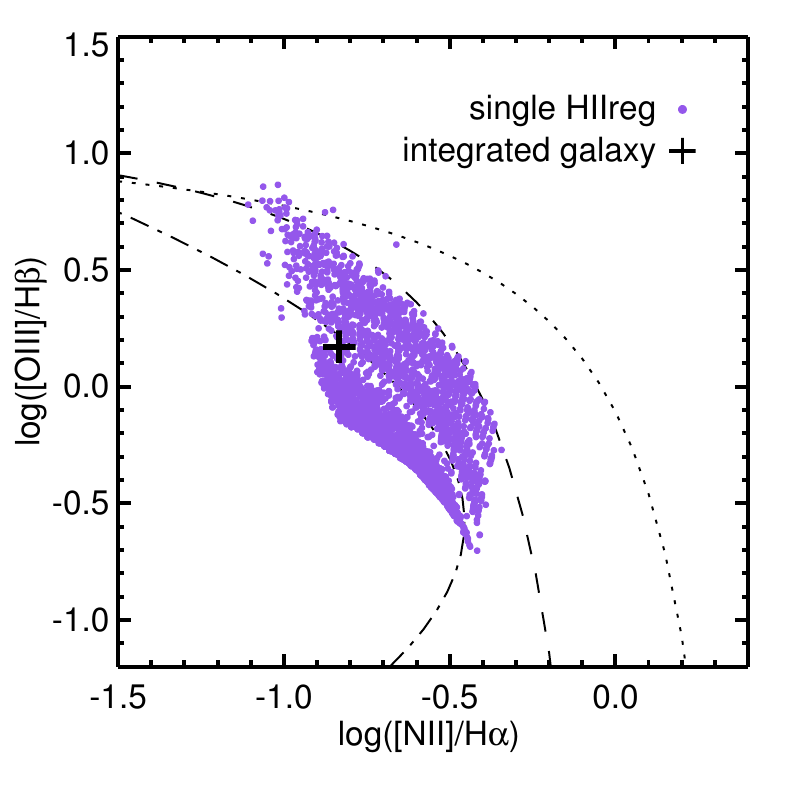}
    \caption{
            \label{BPT_SFR1_fiducial} 
    Result of the interpolation plotted on the BPT plot for one snapshot of a simulated galaxy. The dashed line is the \cite{kauffmann03} line, an empirical separation of star forming galaxies and AGN based on the SDSS galaxies. The dotted line is the \cite{kewley01} theoretical limit for star forming galaxies. The dot-dashed line is a fit to the star forming galaxies from the SDSS galaxies \cite{brinchmann08}. Each \HII region is plotted with a dot (as in Figure \ref{RIF_S_SFR1_fiducial}) and the integrated galaxy is shown with the plus sign. }

  \end{figure}

Figure \ref{RIF_S_SFR1_fiducial} illustrates the procedure. The Figure shows the ionization front radii ($R_{\rm IF}$) and ionizing luminosities ($S$) of all the \HII regions present at a single time step in one of our population synthesis calculations, overlaid with a grid of models for single \HII regions at an age of 0 Myr. The model grid is characterized by values of density $n_{\rm H, in}$ at the inner edge of the \HII region and by the ratio of the ionization front radius to the characteristic radius $\rch$, defined by \citet{yeh12} as the value of $R_{\rm IF}$ for which gas pressure and unattenuated radiation pressure at the ionization front are equal. This radius is given by
\begin{equation}
\label{eq:rch}
\rch = \frac{\alpha_B L^2}{12\pi (2.2 k_B T_{\rm II} c)^2 S},
\end{equation} 
where $k_B$ is the Boltzmann constant, $L$ is the bolometric luminosity, $T_{\rm II}= 7000$ K is the temperature of the ionized gas and the factor 2.2 is obtained by assuming that helium is singly ionized everywhere. Since $L=\psi S \epsilon_0$, the value of $\rch$ is simply proportional to $S$. For the simplest case of \HII regions with an age of 0 Myr, we assign each one a luminosity in the [\OIII],  [\NII], H${\alpha}$ and H${\beta}$ lines by interpolating between the line luminosities of the nearest points in the overlaid model grid. The procedure for older \HII regions is analogous, except that there is an additional interpolation in age. Once we have assigned a luminosity to each \HII region, the total line luminosity of the galaxy is simply the sum over individual \HII regions. Figure \ref{BPT_SFR1_fiducial} shows the final result, where we have used the computed line ratios of both the individual \HII regions from Figure \ref{RIF_S_SFR1_fiducial} and the integrated galaxy to place them in the BPT diagram.

\section{Results}\label{S:results}

Our aim is to investigate how radiation pressure and stellar winds affect galaxies' emission line ratios. As discussed above, the effects are both internal -- changing the density distribution and thus the emission produced within single \HII regions -- and external -- changing the distribution of \HII region radii and other properties. It is easiest to understand the results if we tackle the internal effects separately first, which we do in Section \ref{sec:results_internal}. Then in Section \ref{sec:results_dynamics} we consider external effects and how these interact with internal ones. In Section \ref{SS:EnvPara} we consider how the results depend on the properties of the galaxy as a whole (e.g.~star formation rate, ambient pressure).

\subsection{Internal effects of radiation pressure and winds}
 \label{sec:results_internal}

We first examine how our four internal structure models from Table \ref{tab:models} distribute \HII regions in the BPT diagram.

  \subsubsection{Models with weak winds}

\begin{figure}
%    \epsscale{0.8}
     \plotone{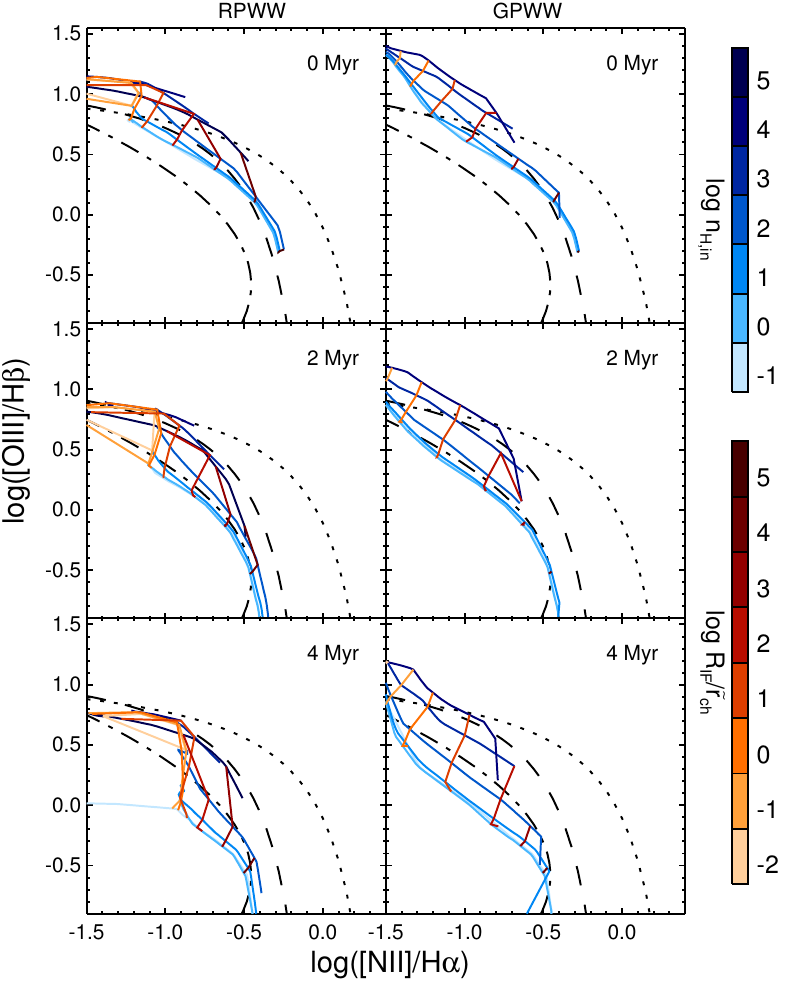}
       \caption{BPT diagram for the models with weak winds evolving from 0 to 4 Myr (from top to bottom). The left column shows Model RPWW (with radiation pressure) and the right column shows Model GPWW (without radiation pressure). The models are shown with lines of constant $\log n_{\rm H, in} = -1$ to 5 (blue) and constant $\log R_{\rm IF}/\rch$ (red), where $R_{\rm IF}$ is the ionization front radius and $\rch$ is the characteristic radius in \cite{yeh12} at which radiation and gas pressure balance.} 
    \label{BPT_Sherry_Model1-3}
  \end{figure}

We compare the two models with weak winds, RPWW and GPWW, in Figure \ref{BPT_Sherry_Model1-3}. We show \HII region models with constant $\log n_{\rm H, in} = -1$ to 5 (blue) and with constant $\log R_{\rm IF}/\rch$ (red), where $R_{\rm IF}$ is the ionization front radius and $\rch$ is the characteristic radius in \cite{krumholz09} at which radiation and gas pressure balance. The ratio $R_{\rm IF}/\rch$ is related to the ionization parameter, as discussed in Paper I. Within each column, we plot three stages of the evolution of the cluster: 0, 2 and 4 Myr (from top to bottom). 

We confirm some trends that have been seen in the past \citep{dopita00, dopita06, kewley01}, such as the decrease of line ratios as the cluster ages and increase of the ionization parameter from bottom right to top left. We explore for the first time a large range of values for the density. We find that the higher the density the stronger the [\NII] and [\OIII] emission, up to the point that the gas density exceeds $\sim 10^4$ cm$^{-3}$. Beyond this, the density in the \HII region exceeds the critical densities of the [\NII] and [\OIII] lines ($6.6\times 10^4$ cm $^{-3}$ and $6.8\times 10^5$ cm$^{-3}$, respectively) causing the line intensity to stop increasing. However, before this point is reached, in the highest density models the [\OIII] emission is large enough that the [\OIII]/H$\beta$ ratio exceeds the upper limit for starburst models described by \cite{kewley01} (black dotted line in Figure \ref{BPT_Sherry_Model1-3}). 

We can understand why our models exceed the \citet{kewley01} limits as follows. \citeauthor{kewley01}~created a grid of photoionization models with fixed initial density n$_{\rm H, in}$ = 350 cm$^{-3}$ and strong stellar winds (i.e.~assuming planar geometry), with a range of metallicities and ionization parameters, and without the effect of radiation pressure. They find that the line ratios in their model never exceed the limit indicated by the black dotted line in Figure \ref{BPT_Sherry_Model1-3}. Our models exceed this limit because they reach regimes of very high density and very high radiation field that the \citeauthor{kewley01}~models, due to their assumption of a fixed density and planar geometry, are unable to access. The underlying physical processes become clear if we compare our various models. Both models GPWW and RPWW can exceed the \citeauthor{kewley01}~limit, while our strong wind models either do not exceed or barely exceed it (see Section \ref{SS:resultSW}). In model GPWW, the density of the gas near the ionizing source can remain unphysical high even when the luminosity is very high; as a result there is significant emission from high-density, highly-irradiated gas. By contrast, in model RPWW, strong radiation pressure pushes gas away from the ionizing source when the luminosity is high, which in turn reduces the amount of gas that is both dense and highly irradiated. This model still breaks the \citeauthor{kewley01}~limit, but by less than GPWW. When stellar winds are included, on the other hand, the wind pushes the gas away from the source, reducing the radiation flux it experiences. This strongly limits the amount of dense, highly-irradiated gas in both of our strong wind models, and in the \citeauthor{kewley01}~models. We therefore see that the \citeauthor{kewley01}~limit is not a limit imposed by the physics of \HII regions in general; instead, it is driven by  \citeauthor{kewley01}'s assumptions about the structure of \HII regions, and the limitations on density and ionizing luminosity that these assumptions imply.

Comparing the cases with and without radiation pressure, we see that models with radiation pressure often produce less [\OIII] emission that those without. This effect arises because \HII regions with radiation pressure and that have $R_{\rm IF}/\rch \ll 1$ have most of their gas in a radiation-confined shell that has a steep density gradient. This should be compared to the mostly uniform density produced if one ignores radiation pressure (\citealt{draine11, yeh12}; Paper I). The higher density in this shell means that the density in the bulk of the emitting gas can exceed the critical density for a line even when the mean density of the \HII region is below this value. Hence, Model RPWW saturates at a lower value of [\OIII]/H$\beta$ than model GPWW.

  \subsubsection{Models with strong winds}\label{SS:resultSW}
  
\begin{figure}
%      \epsscale{0.8}
\plotone{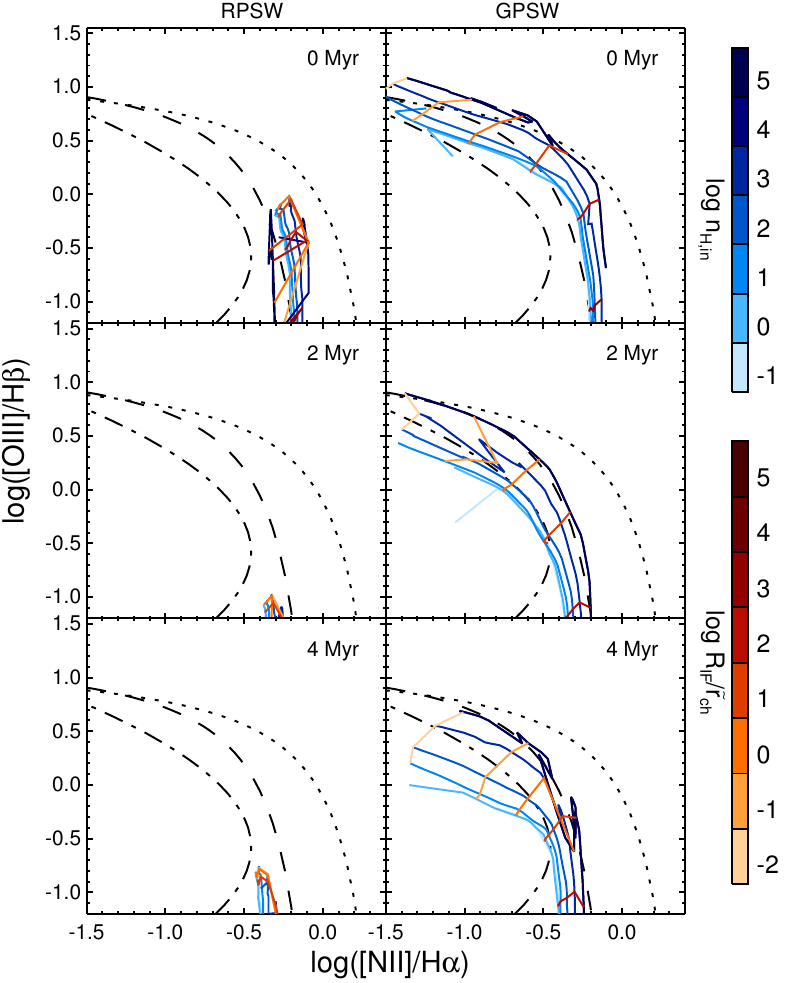}
    \caption{
    \label{BPT_Sherry_Model2-4}
    Same as Figure \ref{BPT_Sherry_Model1-3}, but for the strong-wind Models GPSW and RPSW.
    }
  \end{figure}

We show the BPT diagram locations of the strong stellar wind models, RPSW and GPSW, in Figure \ref{BPT_Sherry_Model2-4}. The first thing that is evident from the Figure is that Model RPSW produces line ratios in the BPT diagram far from both the other models and from the locations of observed galaxies. The region of parameter space where the models are not physical within the context of RPSW corresponds to \HII regions with large ionizing luminosities but small radii, and one can understand why Model RPSW avoids this region with a small thought experiment. A value of $\Omega=100$ implies that $P_{\rm IF} V_{\rm IF}  / P_{\rm in} V_{\rm in} - 1 \ll 1$ (see Equation~\ref{EQ:Omega}), meaning that the shocked wind gas dominates the total energy budget. This in turn requires that  $V_{\rm IF} \approx V_{\rm in}$ and $P_{\rm IF} \approx P_{\rm in}$, so that the wind bubble fills almost the entire volume of the \HII region, leaving only a thin shell of photoionized gas, and the gas pressures are nearly identical at the inner and outer edges of this shell. However, if the ionizing luminosity is large enough (specifically if it is large enough so that $R_{\rm IF} < \rch$), this is impossible. As $S\rightarrow \infty$ the radiation pressure at the inner edge of the photoionized shell must greatly exceed the gas pressure, and the gas pressure $P_{\rm IF}$ at the outer edge of the \HII region, where all of the radiation has been absorbed, must be equal to the \textit{total} pressure at the inner edge, which is the sum of the small gas pressure $P_{\rm in}$ and the much larger radiation pressure. It therefore follows that at sufficiently large $S$ one must have $P_{\rm IF}/P_{\rm rad} \gg 1$, giving $\Omega \ll 1$, a point also made by \citet{yeh12}. Thus one cannot simultaneously have arbitrarily large $S$, arbitrarily small $R_{\rm IF}$, and $\Omega \gg 1$. This issue is discussed further in Paper I.

This problem does not affect Model GPSW, since in this model one ignores radiation pressure. These models thus represent wind-dominated \HII regions, and are qualitatively similar to the models of \cite{dopita00} and \cite{kewley01}. In Paper I, we show a comparison of these models with those of \cite{dopita00}, and find a good match with their results.

  \subsection{Dynamical effects of radiation pressure}\label{sec:results_dynamics}

Having understood the effects of radiation pressure and winds on the internal structure of \HII regions, we are now ready to study their dynamical effects.

\subsubsection{Distribution of \HII region radii}

\begin{figure}
%    \epsscale{0.6}
      \plotone{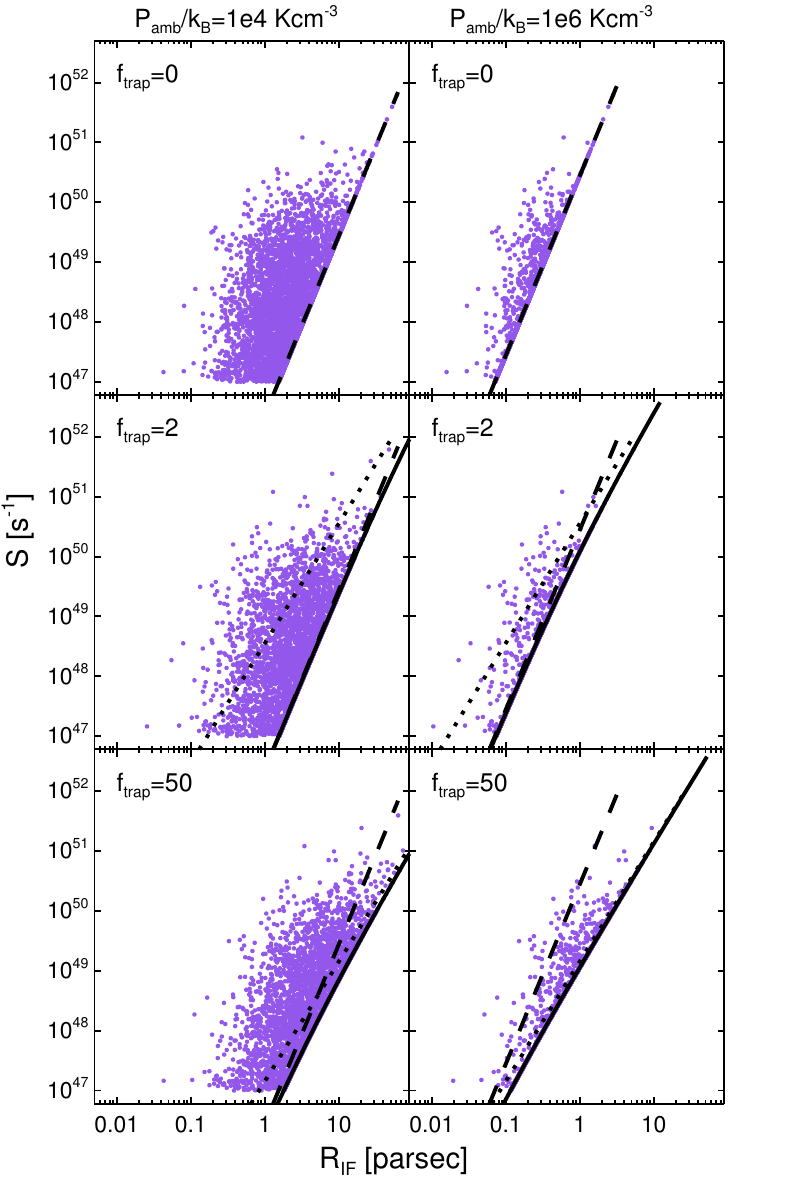}
      \caption{
            \label{RIF_S_Model1}
            Radius of the ionization front $R_{\rm IF}$ versus ionizing photon luminosity $S$ for all the \HII regions present at one time step in simulated galaxy (dots). We show simulations with two different values of $P_{\rm amb}/k_B$ ($10^4$ K cm$^{-3}$ in the left column, and $10^6$ K cm$^{-3}$ in the right column), and three different values of $f_{\rm trap}$ (0 in the top row, 2 in the middle row, 50 in the bottom row). Black lines show the location of the stall radii for the simulations, with dashed lines corresponding to stalling when the pressure is gas-dominated, dotted lines to stalling when the pressure is radiation-dominated and full lines when to stalling when both radiation and gas pressure are relevant. The SFR is 1 $M_{\sun}$ yr$^{-1}$ in all the simulations shown, so the number of \HII regions present in each panel is approximately the same. }
  \end{figure}

  \begin{figure*}
%      \epsscale{1}
      \plotone{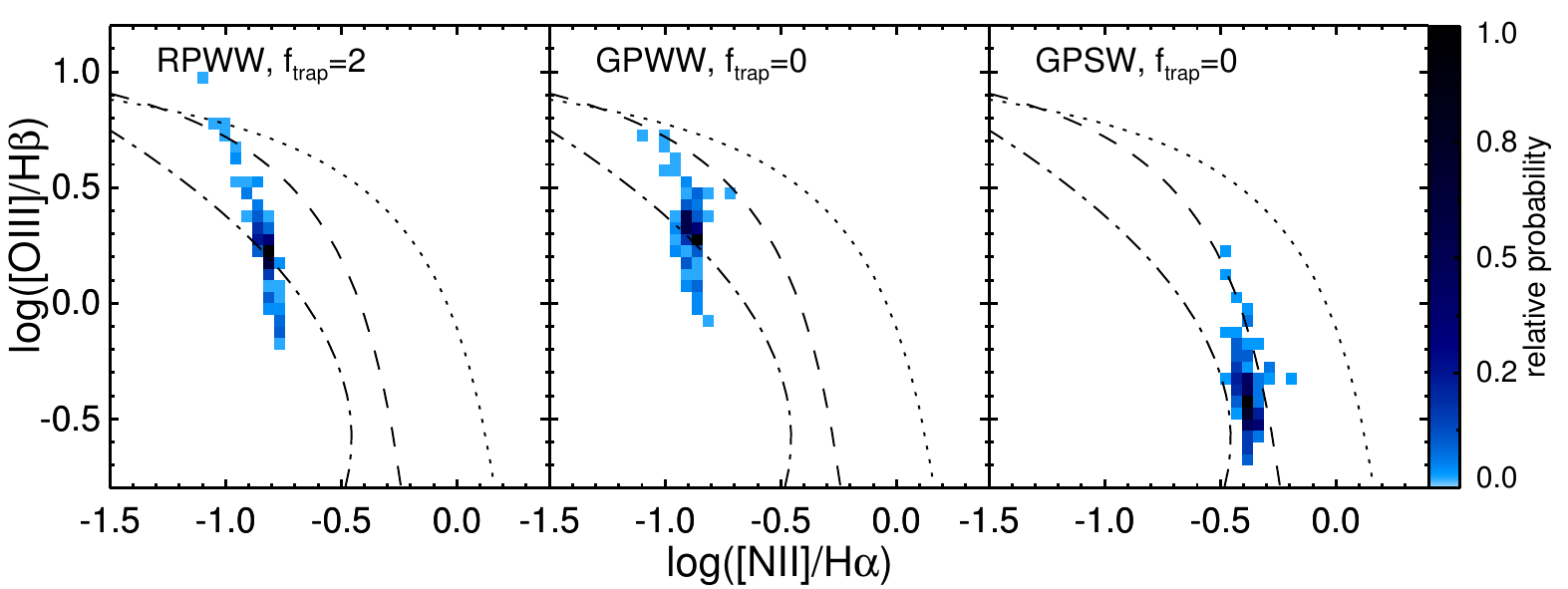} 
      \caption{Synthetic galaxies on the BPT diagram generated with our fiducial parameters and \HII region modeling. Each model is a single time snapshot from our simulations, with the snapshots taken at intervals of 1 Myr. The region shown has been rasterized into pixels of (0.05 dex)$^2$. The color in each pixel corresponds to the number of models that fall into that pixel, normalized by the pixel containing the most models. The three cases shown are Model RPWW with $f_{\rm trap} = 2$, corresponding to \HII regions with weak winds whose dynamics and internal structure include radiation pressure; Model GPWW with $f_{\rm trap} = 0$, corresponding to \HII regions that are classical Str\"{o}mgren spheres with neither radiation nor wind pressure affecting their internal structure or dynamics; and Model GPSW with $f_{\rm trap} = 0$, for which \HII regions are wind-dominated bubbles without radiation pressure.}
      \label{BPT_fid_compareModels}
      
  \end{figure*}

In the expansion of an \HII region, the radiation pressure term contributes as an additional push towards a faster radial expansion. To study this effect we examine the distribution of \HII region radii produced by our population synthesis code, and how it is influenced by radiation pressure. Figure \ref{RIF_S_Model1} shows a scatter plot of the radius of the ionization front ($R_{\rm IF}$) versus the ionizing photon luminosity ($S$) for all the \HII regions present at one time step in two of our simulations, one with $P_{\rm amb}/k_B=10^4$ K cm$^{-3}$ (left column) and one with $P_{\rm amb}/k_B = 10^6$ K cm$^{-3}$ (right column). We show three cases: $f_{\rm trap}=0$ is a model where radiation pressure does not affect the dynamics at all, $f_{\rm trap}=2$ is our fiducial case, and $f_{\rm trap} = 50$ is a model where the radiation pressure is assumed to be strongly trapped within the \HII region, and affects the dynamics much more strongly. The case $f_{\rm trap} = 0$ corresponds to \HII regions that follow the classical \citet{spitzer78} solution, $f_{\rm trap} = 2$ corresponds roughly to the value favored by the radiation-hydrodynamic simulations of \citet{krumholz12c, krumholz12f}, while $f_{\rm trap} = 50$ corresponds to the peak of the values adopted in the subgrid models of \citet{hopkins11a}, where radiation is assumed to build up inside \HII regions and produce large forces. In each panel we also show with full lines the stalling radii, defined as the radii where the internal pressure of the \HII region drops to $P_{\rm amb}$. Each \HII region, when is created, is assigned a value of $S$ and has $R_{\rm IF}=0$. As time passes, the \HII region evolves and moves horizontally in the $R_{\rm IF}$ versus $S$ plane till it reaches this limiting line at the stall radius. Since $P_{\rm rad}/P_{\rm gas}$ decreases as $R_{\rm IF}$ grows at fixed $S$, depending on the value of $S$ and $P_{\rm amb}$, this can happen when $P_{\rm gas} \ll P_{\rm rad}$, when $P_{\rm gas} \gg P_{\rm rad}$, or when $P_{\rm gas} \simeq P_{\rm rad}$. If the \HII region stalls when the gas is dominated by radiation pressure, $P_{\rm amb} \simeq P_{\rm rad}$, and from equation (\ref{EQ:Prad}) we have $R_{\rm IF} \propto S^{1/2}$; if stalling occurs when an \HII region is dominated by gas pressure, then $P_{\rm amb} \simeq P_{\rm gas}$, and from equation (\ref{EQ:Pgas}) we have $R_{\rm IF} \propto S^{1/3}$. Figure \ref{RIF_S_Model1} shows also these two dependencies as dotted and dashed lines respectively.

The Figure shows that radiation pressure has two distinct effects on the dynamics. First,  \HII regions with radiation pressure expand faster than classical ones, so that models are shifted to increasingly large values of $R_{\rm IF}$ as $f_{\rm trap}$ increases. The shift from $f_{\rm trap} = 0$ to 2 is relatively modest, while the gap between $f_{\rm trap} = 2$ and 50 is somewhat larger, corresponding to nearly half a dex in radius. The second effect of radiation pressure is to increase the stalling radius. When the ambient pressure is small, this has a relatively small effect, because the stalling radius is large and most \HII regions turn off before reaching it. One the other hand, when the pressure is high, the stalling radius is smaller and most \HII regions stall before their driving stars evolve off the main sequence. In this case most \HII regions are clustered up against the stalling radius, and the increase in stalling radius with $f_{\rm trap}$ has very significant effects. 

\subsubsection{Distribution of \HII regions in the BPT diagram}

We are now ready to use our population synthesis code to determine where simulated galaxies lie in the BPT diagram. We run three classes of models. The first, which we consider the most physically realistic given the observed properties of \HII regions in the local Universe, uses Model RPWW for the internal structures of \HII regions, and uses $f_{\rm trap} = 2$ to determine their dynamical evolution. The other two models use $f_{\rm trap} = 0$ (i.e.~assume that \HII regions expand as classical Spitzer \HII regions), and use Models GPWW and GPSW for the internal structures. The latter choice is not fully consistent, in that with strong wind models we should use a wind-dominated dynamical solution such as that of \citet{castor75a}. We do not do so, however, both because \citet{dopita06} have already explored this case, and because observations now strongly disfavor it.

Figure  \ref{BPT_fid_compareModels} shows the comparison of the three models on the BPT diagram for our fiducial parameter choices (see Table \ref{tab:fiducial}). Each model represents the line ratio produced by summing the line emission over all the \HII regions present in a simulated galaxy at a given snapshot in time, and for each model we show 200 such snapshots, separated by intervals of 1 Myr. The region shown in the plot has been rasterized into pixels of (0.05 dex)$^2$. The color in each pixel corresponds to the number of models that fall into that pixel, normalized by the pixel containing the most models. The plot shows several interesting results. Model GPSW, in which \HII regions' internal structures are wind-dominated, are systematically shifted to lower [\OIII]/H$\beta$ and higher [\NII]/h$\alpha$ than the weak wind models. Model RPWW spans a wide range of parameter space, including some snapshots that exceed the \citet{kewley01} theoretical limit. These snapshots tend to be immediately after the formation of a very large, bright, association. Model GPWW is covers a smaller range in the plot, and stays below the \cite{kewley01} limit.

\begin{figure}
      %\epsscale{0.8}
\plotone{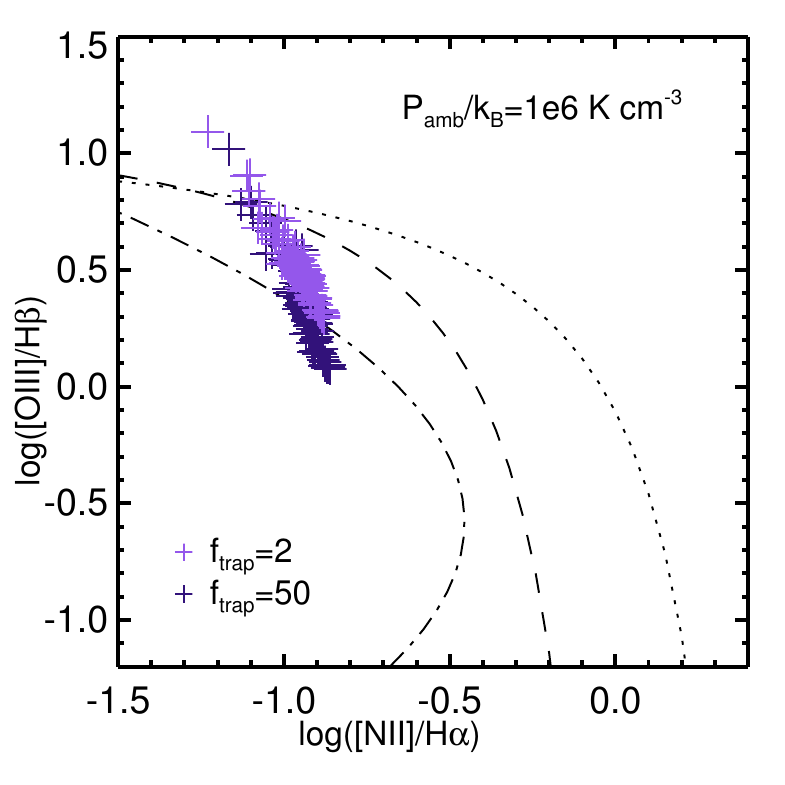}
\caption{
\label{fig:ftrapcomp}
Comparison of BPT diagrams for two runs with $P_{\rm amb}/k_B = 10^6$ K cm$^{-3}$ and $f_{\rm trap}=2$ and 50. Each plus sign represents a single time snapshot from our simulations, with the snapshots taken at intervals of 1 Myr. Both runs use Model RPWW, and are otherwise identical to the runs shown in Figure \ref{BPT_fid_compareModels}. 
}
\end{figure}

We varied a number of the fiducial parameters, and found them to have little effect on the results. Parameters whose influence is negligible include $M_{a,\rm min}$ and $M_{a,\rm max}$, minimum and maximum value of the association mass, $\mathcal{M}$, the Mach number used to set the width of the density distribution, and $k_{\rm \rho}$ the powerlaw index that describes the density distribution into which \HII regions expand. Perhaps surprisingly, the value of $f_{\rm trap}$ also has relatively little effect if we hold the internal models fixed, as illustrated in Figure \ref{fig:ftrapcomp}. In other words, if we use Model RPWW to describe the internal structure of \HII regions, the differences in the distributions of \HII region radii visible in Figure \ref{RIF_S_Model1} as we vary $f_{\rm trap}$ from 0 to 50 do not produce corresponding differences in the locations of the resulting galaxies in the BPT diagram -- or at least the differences they produce are mostly within the scatter produced simply by stochastic drawing of association masses and surrounding densities. In Figure \ref{RIF_S_SFR1_fiducial} we show the grid of models covering over 5 orders of magnitude both in $R_{\rm IF}$ and S. The grid dramatically shrinks in the BPT diagram (Figure \ref{BPT_Sherry_Model1-3}) causing the small effect of $f_{\rm trap}$ in Figure \ref{fig:ftrapcomp}. Thus there does not appear to be an obvious way to use line ratio observations of integrated galaxies to measure the value of 
the dynamical parameter $f_{\rm trap}$. We stress, however, that $f_{\rm trap}$ includes the influence of wind pressure, and a wind-pressure dominated state can be identified, on the basis of line ratio observations, through its effect on the internal structures of \HII regions.  In particular, wind-dominated regions cannot access high values of the ionization parameter and are limited to the lower right of the BPT diagram; see Paper I and \citet{yeh12} for a thorough discussion.

  \subsection{Influence of galactic parameters}\label{SS:EnvPara}

  \begin{figure*}
      \centering
%      \epsscale{1}
      \plotone{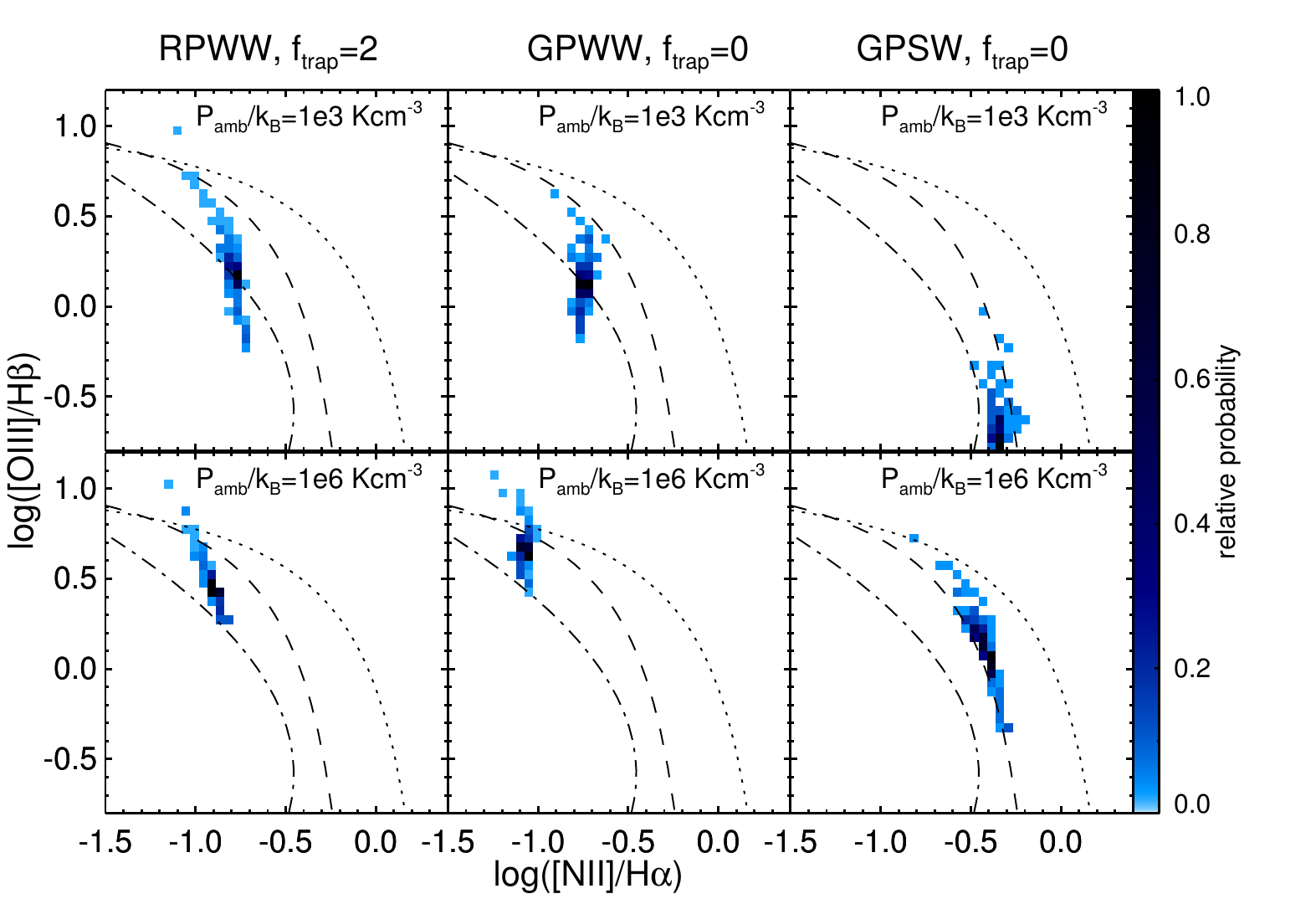} 
      \caption{Simulation results with varying $P_{\rm amb}$. Models RPWW, GPWW, and GPSW are plotted respectively on the left, center and right column for $P_{\rm amb}/k_B = 10^3$ and $10^6$ K cm$^{-3}$ (top and bottom rows). The region shown has been rasterized into pixels of (0.05 dex)$^2$. The color in each pixel corresponds to the number of models that fall into that pixel, normalized by the pixel containing the most models. All other parameters are the same as in the fiducial case.} 
      \label{BPT_Pamb_SFR1}
  \end{figure*}

  While there are a number of parameters in our model that make very little difference to the results, the two parameters $P_{\rm amb}$ and SFR that we use to characterize our galaxies do have a measurable influence. Figure  \ref{BPT_Pamb_SFR1} shows how the ambient pressure influences the position of simulated galaxies on the BPT diagram. We show our three models computed with $P_{\rm amb}/k_B = 10^3$ and $10^6$ K cm$^{-3}$ (top and bottom rows). At low ambient pressure, we find a significantly larger spread in the simulated galaxies. This is because for low ambient pressure the stalling radius is large, many \HII regions do not live to reach it, and thus \HII regions span a large range of radii. Exactly where \HII regions fall in the plane of $S$ and $R_{\rm IF}$ is therefore subject to a great deal of stochastic variation. In contrast, as show in Figure \ref{RIF_S_Model1}, increasing the ambient pressure causes all the \HII regions in a galaxy to cluster along the stall radius line. In Figure  \ref{BPT_Pamb_SFR1} we can also see that the ambient pressure controls the overall location in the BPT plot, moving all the synthetic galaxies to a higher position in the BPT diagram, and at higher ionization parameter.

    \begin{figure*}
      \centering
      \plotone{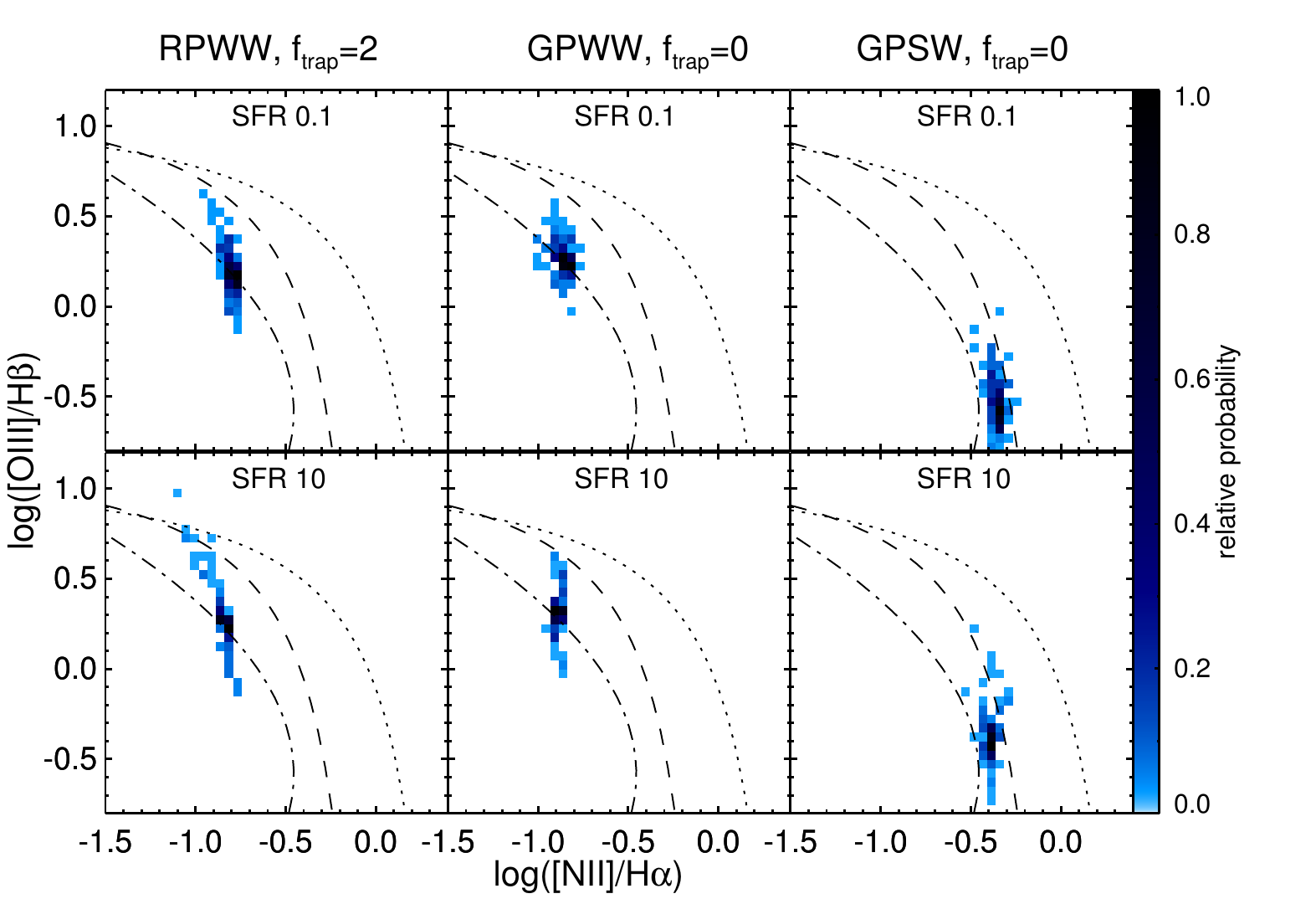} 
      \caption{Same as Figure \ref{BPT_Pamb_SFR1}, but here the top and bottom rows correspond to star formation rates of $0.1$ and $10$ $M_{\odot}$ yr$^{-1}$. All other parameters are the same as in the fiducial case. 
      } 
      \label{BPT_Pamb1e4_SFR}
  \end{figure*}

Figure \ref{BPT_Pamb1e4_SFR} shows the effects of varying the SFR on the location of our synthetic galaxies on the BPT diagram. As the Figure shows, a smaller SFR leads a bigger spread of points in the BPT diagram. This is due to the stochastic nature of star formation at low SFRs, something that can also lead to large variations in absolute line fluxes as well as line ratios \citep{fumagalli11a, da-silva12a, weisz12a}. If we draw a large mass for the next association to be created, a long time passes until it appears, especially when the SFR is low. During this phase there are no young, bright \HII regions present, and so the galaxy is located in the bottom-right part of the BPT plot. When the association finally forms, the galaxy's line emission becomes dominated by the resulting bright, young \HII region, which drives it to the top-left part of the BPT diagram. As a result, there is a great deal of variation in the galaxy's location. When the SFR is high, on the other hand, \HII regions form continuously, causing the population of \HII regions to be more numerous and uniform. 
We do caution that our mechanism for handling \HII region creation may overestimate the amount of stochasticity found in real galaxies, but that the general sense of the effect will be the same as we have found, even if its magnitude is overestimated. A more realistic formalism for handling the problem of drawing association masses and birth times subject to an overall constraint on the star formation rate is implemented in the SLUG code \citep{da-silva12a}; adding this formalism to our code is left for future work.

  \section{Comparison to observations} \label{S:obs}
  
Having understood the physics that drives the location of galaxies in the BPT diagram, we are now in a position to compare our models to observations. Such observations come in two varieties: spatially resolved ones of individual \HII regions or portions of galaxies, and unresolved ones in which the line fluxes from all the \HII regions in a galaxy are summed. Since our code produces collections of stochastically-sampled \HII regions, we can compare to both.
For reference and to facilitate comparison, we show in both cases unresolved observations of local galaxies from the SDSS \citep{brinchmann04, tremonti04} along with a fit to this sequence \citep{brinchmann08}, the empirically-determined line separating star-forming galaxies from AGN \citep{kauffmann03}, and the  \cite{kewley01} theoretical upper limit to star forming galaxies. The single \HII region sequence and the SDSS star forming galaxy sequence overlap, at least in the upper left part of the BPT diagram, while high redshift galaxies seem to create a different sequence, upward and to the right \citep{liu08,  brinchmann08, hainline09, erb10}. 

\begin{figure}
  \centering
  %\epsscale{0.8}
  \plotone{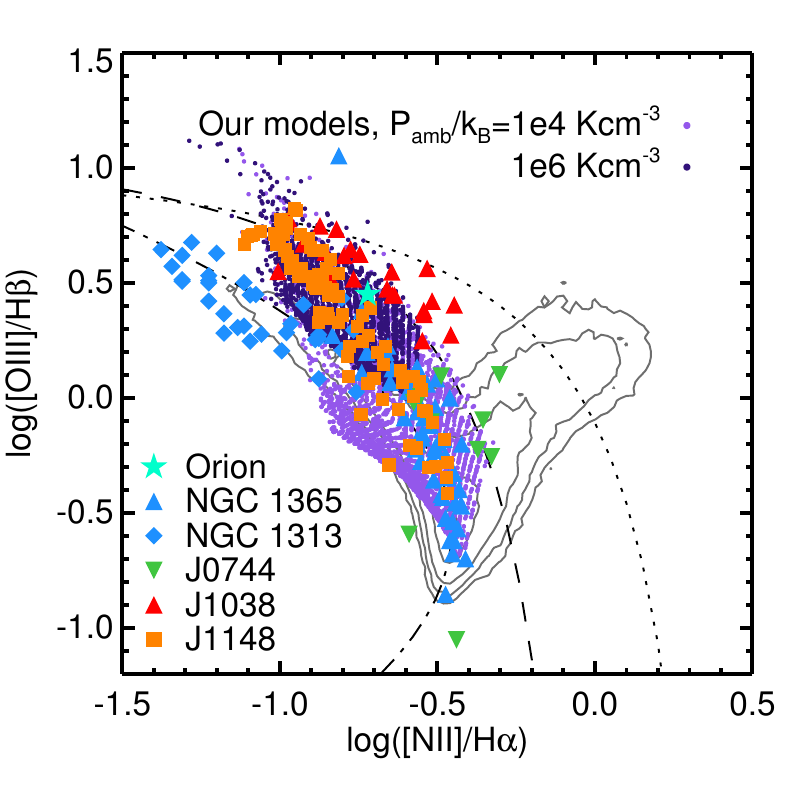} 
  \caption{A comparison between resolved \HII regions and individual \HII regions produced in our simulations in the BPT diagram. We show \HII regions in galaxies at $z=0$ with blue symbols; the galaxies shown are NGC 1365 \citep{roy97} (triangles), NGC 1313 \citep{walsh97} (diamonds), and the integrated value for the Orion nebula \citep{sanchez07} (star). We also show single spatial pixels measured with OSIRIS for three lensed galaxies at redshift $z\sim2.0-2.4$ from \citep{jones12} (green, red, and orange). Contours represent galaxies from the SDSS catalogue \citep{brinchmann04}, enclosing respectively, 5, 10, 20, 50, 90 and 99\% of all galaxies in which the four emission lines are detected at a greater than 3$\sigma$ significance level. Blue and purple points show the results of our models using fiducial parameters and $P_{\rm amb}/k_B = 10^4$ and $10^6$ K cm$^{-3}$, respectively. Finally, the dotted line is the theoretical upper limit of \citet{kewley01}, the dashed line is the empirical AGN - star-forming galaxy separating line \citep{kauffmann03}, and the dot-dashed line is the fit to the SDSS sample of \citet{brinchmann08}.
  \label{obs1}
  }
\end{figure}

Figure  \ref{obs1} shows a collection of observations of single \HII regions and pixel by pixel observations taken from the literature. For the local Universe, our comparison data set consists of single \HII regions from NGC 1365 \citep{roy97}, NGC 1313 \citep{walsh97}, and the Orion region in our own galaxy \citep{sanchez07}. We also plot individual pixels in three lensed galaxy at $z\sim 2$ from \citet{jones12}, which scatter about a locus that passes close to the location of Orion in the BPT diagram. As pointed out by \citet{walter09}, the SFR surface density of Orion is similar to that of a high redshift object undergoing a burst of star formation. 

On top of these data, we overlay the results of our simulations using model RPWW, which we consider most realistic based on observations of nearby \HII regions. The results shown are single snapshots of all the \HII regions produced in two different simulations, one with $P_{\rm amb}/k_B = 10^4$ K cm$^{-3}$ one with $10^6$ K cm$^{-3}$. These two cases should roughly bracket what we expect for Milky Way-like galaxies and for the dense, more strongly star-forming galaxies found at high redshift. The plot shows that our simulations are able to roughly reproduce the locus of observed \HII regions in the BPT diagram for a reasonable range of ambient pressures. We cannot reproduce most of the \HII regions in NGC 1313, because the galaxy has a metallicity lower than solar and our model considers only solar metallicities. Lower metallicity produces a shift of the models towards lower [\NII]/H${\alpha}$ values \citep{dopita00}. The pixel by pixel high-$z$ galaxies are best fit by the models with high $P_{\rm amb}$, consistent with observations that these galaxies have high surface and volume densities.

\begin{figure}
  \centering
   %\epsscale{0.8}
  \plotone{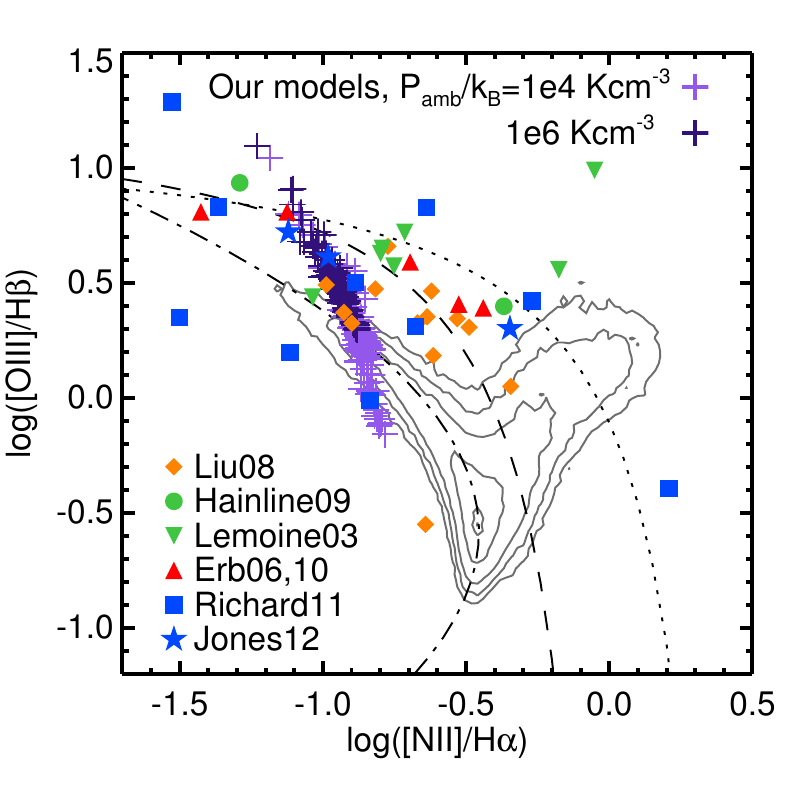} 
  \caption{A comparison between observed unresolved galaxies and simulated produced by our code in the BPT diagram. Contours and lines are the same as in Figure \ref{obs1}. Diamonds show galaxies at $z\sim 1- 1.4$ taken from the DEEP2 survey \citet{shapley05, liu08}; triangles show galaxies at $z\sim 2$ taken from \citet{erb06a, erb10}; lensed galaxies at a variety of redshifts and samples are indicated by inverted triangles \citep{lemoine03}, circles \citep{hainline09}, squares \citep{richard11}, and stars \citep{jones12}. Blue and purple plus signs show the results for integrated galaxies in of our models using fiducial parameters, Model RPWW, and ambient pressures of $P_{\rm amb}/k_B = 10^4$ and $10^6$ K cm$^{-3}$. Each point represents a single time snapshot taken at 1 Myr intervals from a simulation that evolves for 200 Myr.
  \label{obs2}
  }
\end{figure}

Figure  \ref{obs2} shows the comparison with integrated galaxy measurements; these come from the SDSS for the local Universe, and from a variety of surveys at high-$z$. Many SDSS star forming galaxies lie in the lower part of the star forming sequence due to the presence of a diffuse warm component in the interstellar medium. \citet{brinchmann04} point out that a significant amount of the emission line flux in these galaxies comes from the diffuse ionized gas, rather than from \HII regions. The combination of the diffuse ionized gas and the \HII regions typically has a lower effective ionization parameter, and compared to \HII regions alone it shows an enhanced [\NII]/H${\alpha}$ and depressed [\OIII]/H${\beta}$ \citep{mathis00}. Therefore, we only expect our models, which do not include the diffuse ionized gas, to reproduce the upper part of the star forming sequence of the SDSS. 

In Figure \ref{obs2} we also overplot the whole-galaxy results produced by our code. As the plot shows, while we are able to reproduce the full spread of individual \HII regions, our simulations of whole galaxies cover a more limited range of BPT than the observations. In particular, we tend to underpredict the observed [\NII]/H$\alpha$ ratios. There are several possible explanations for why we might successfully reproduce individual \HII regions, even in high-$z$ galaxies, but not fully cover the range of integrated galaxy properties. One we have already discussed in the introduction: the offset at high-$z$ may be due to the contribution of a weak AGN, which our models obviously do not include. A second possibility is that the contribution of diffuse ionized gas to the line ratios cannot be neglected even in these high redshift galaxies. Another is that our weighting of the different \HII regions is incorrect because the association mass function is different than the $-2$ powerlaw we have adopted based on local observations, or because of biases introduced by dust extinction, despite the extinction-independent nature of the BPT line ratios (see \citealt{yeh12}). A fourth possibility is that our lognormal distribution of densities provides a poor fit to the true range of densities into which \HII regions expand in high-$z$ galaxies, so that the amount of time individual \HII regions spend in the upper left versus the lower right parts of the BPT diagram is off in our models.

As a last possibility, we recognize that the ability of our models for individual \HII regions to reproduce the observations of Orion very well (Figure \ref{obs1}) may be partly a matter of good luck. Our models are not designed to mimic the champagne flow phase of young ($<10^5$ yr), compact \HII regions. In particular, we assume a state of quasi-static force balance which holds only approximately in accelerating flows; see \citet{yeh12} \S~3.4 on this point. 
 Indeed, Orion does not resemble the typical \HII region - e.g., a few million years old and at the stalling radius - in our galaxy simulations. It is possible that the different distribution of high-$z$ galaxies in the BPT plot as compared to local SDSS galaxies is due to the higher pressure environment in the former, which keeps the \HII regions longer in a champagne flow-like phase. 
  Future studies might assess our models' accuracy in the champagne phase, extend their range of validity, and quantify the importance of this dynamical detail for high-$z$ galaxies.

\section{Discussion and Conclusions}\label{S:conclusions}

Motivated by recent observations suggesting that \HII regions are shaped much less than expected by the pressure of shocked stellar wind gas, and much more by direct radiation pressure \citep{harperclark09, lopez11, yeh12}, we revisit the problem of determining the line flux emitted by a population of \HII regions. We adopt as our default a model of \HII regions where the pressure of winds is subdominant, and radiation pressure is not neglected, and we compare this result to traditional models with strong winds and weak radiation pressure. In Paper I we discuss how we generate grids of static, single \HII regions, with a wide range of sizes and ionizing luminosities, with varying strengths of winds and radiation pressure. In this paper we construct dynamical expansion models for these \HII regions, and explore how changing the strength of winds and radiation pressure affects their line ratios in the BPT diagram. We find that radiation pressure has two important effects. First, \prad\ changes the internal structure of the \HII region, creating a density gradient towards the outer shell. This affects the expected line emission, allowing the \HII regions to exceed the upper limit form starburst models set by \cite{kewley01}. Second, radiation pressure provides an extra boost to the expansion, leading to larger radii at earlier times.

We embed these models in a population synthesis code that generates galactic collections of stochastically-generated \HII regions expanding into a turbulent medium. The code follows \HII regions as they are born, evolve, stall and die. Using this code we predict the integrated line emission of galaxies as a function of several galactic properties. We find that the two most important ones in controlling where galaxies appear in the BPT diagram are the ambient pressure, which shifts galaxies up and to the left as it increases, and the star formation rate, which affects the amount of stochastic scatter in a galaxy's line ratios.

We compare with observations in two distinct ways. First, we select single \HII regions observed in the local universe and pixel by pixel observations of $z\sim 2$ galaxies, and we compare these to the distributions of individual \HII regions produced in our model. We show that our model produces good agreement with the observations for reasonable ranges of SFR and ambient pressure. The high redshift pixel data are best reproduced by \HII regions evolving in a high pressure medium and with high SFR, which we interpret as a sign of intense star formation in a dense interstellar medium, consistent with the observed properties of high-$z$ galaxies \citep[e.g.][]{elmegreen09c, genzel11a}.

Second, we compare integrated galaxies from the SDSS catalogue and the high redshift universe to our synthetic galaxies. We find that, while we are able to reproduce the spread of individual \HII regions, our models for integrated galaxies cluster too tightly compared to the observed range of line ratios in real galaxies, particularly at high-$z$. This might be due to a number of factors. One possibility is that the lognormal distribution of the ambient density we have adopted is a poor description of the density distribution in high-$z$ galactic disks. Another possibility is that winds might be important at high redshift or that the presence of the diffuse ionized medium is not negligible. A third possibility is that a higher pressure environment in high-z galaxies keeps the \HII regions longer in a champagne flow-like phase.  One last possibility is that $z \sim 2$ star forming galaxies may contain an AGN that partially contributes to the line emission.
We leave these possibilities as a subject for future work.

\acknowledgements We thank Jarle Brinchmann, Brent Groves, and Alice Shapley for helpful discussions. This project was initiated during the 2010 International Summer Institute for Modeling in Astrophysics (ISIMA) summer program, whose support is gratefully acknowledged. MRK acknowledges support from an Alfred P.~Sloan Fellowship, from the National Science Foundation through grant CAREER-0955300, and from NASA through Astrophysics Theory and Fundamental Physics grant NNX09AK31G and a Chandra Telescope Grant. 
SCCY and CDM would like to acknowledge an NSERC Discovery grant and conversations with Stephen Ro and Shelley Wright.

\bibliographystyle{apj}

\end{document}